\documentclass{article}
\usepackage{epsfig}
\usepackage{amssymb}

\def\eps{\epsilon}
\def\l{\lambda}
\def\be{\begin{equation}}
\def\ee{\end{equation}}
\def\ba{\begin{eqnarray}}
\def\ea{\end{eqnarray}}

\def\S{{\cal S}}

\def\l{\lambda}
\def\b#1{{\mathbb #1}}

\newcommand{\T}{\mbox{Tr}}
\begin{document}
February 2001         \hfill
\vskip -0.55cm 
\hfill    UCB-PTH-01/06 
 
\hfill  LBNL-47482 
\begin{center}

\vskip .15in

\renewcommand{\thefootnote}{\fnsymbol{footnote}}
{\large \bf Some remarks on unilateral matrix equations}
\footnote{Talk given by the first author at the Euroconference
``Brane new world and noncommutative geometry'', Villa Gualino, Torino, Italy,
October 2-7, 2000}
\vskip .25in
Bianca L. Cerchiai\footnote{email address: BLCerchiai@lbl.gov} and 
Bruno Zumino\footnote{email address: zumino@thsrv.lbl.gov}
\vskip .25in

{\em    Department of Physics  \\
        University of California   \\
                                and     \\
        Theoretical Physics Group   \\
        Lawrence Berkeley National Laboratory  \\
        University of California   \\
        Berkeley, California 94720}
\end{center}
\vskip .25in

\begin{abstract}
We briefly review the results of our paper \cite{CeZu}: we
study certain perturbative solutions of left-unilateral matrix
equations. These are algebraic equations where the coefficients and the
unknown are square matrices of the same order, or, more abstractly, elements 
of an associative, but possibly noncommutative algebra, and all coefficients 
are on the left. Recently such equations have appeared in a discussion of 
generalized Born-Infeld theories. In particular, two equations, their 
perturbative solutions and the relation between them are 
studied, applying a unified approach based on the generalized Bezout 
theorem for matrix polynomials.

\end{abstract}

\section{Introduction}
\setcounter{footnote}{0}

Left-unilateral matrix equations are algebraic equations of the form
\be
P(x)=0, \mbox{ with }  P(x) \equiv a_0+a_1 x+a_2 x^2+\ldots +a_n x^n,
\ee
where the coefficients $a_r$ and the unknown $x$ are square matrices of the
same order and all coefficients are on the left.

The motivation for studying these kinds of equations is that
recently they have appeared in the context of 
generalized Born-Infeld theories~\cite{BMZ,ABMZ1}. The construction
of a self-dual Lagrangian can be reduced to their solution.

We have proposed in \cite{CeZu} a unified approach to these equations
based on the generalized Bezout theorem for matrix polynomials. This
enables us to combine the idea of constructing the trace of a perturbative 
solution in terms of contour integrals in the complex plane of the trace of the
resolvent of the corresponding matrix, due to A. Schwarz~\cite{Schwarz},
with the idea of applying the basic property of the logarithm, as
proposed in \cite{ABMZ2}.

If we define the characteristic polynomial associated to $P(x)$ as
\be
P(\l)=a_0+a_1 \l+a_2 \l^2+\ldots+a_n \l^n, \qquad \l \in \b{C}
\label{characteristic}
\ee
then the statement of the generalized Bezout theorem~\cite{Gant} is that
$\l-x$ is a divisor of $P(\l)-P(x)$ on the right, i.e. it is possible to find
a polynomial $Q(\lambda,x)$, such that
\be
P(\l)-P(x)=Q(\lambda,x)(\lambda-x).
\ee
In fact, in this particular case it is easy to see that
\be
Q(\lambda,x)=\sum_{l=0}^{n-1}\lambda^l \left(\sum_{r=l+1}^n a_r 
x^{r-l-1}\right)\: .
\ee

Notice that all the results described in this paper remain valid in a
more general setting, if one considers $a_r$ and $x$ as elements of 
an associative, but possibly noncommutative, algebra, and
an appropriate algebraic definition of the trace as cyclic average
(see \cite{ABMZ1}) is used.

\section{Properties of the Trace of Perturbative Solutions}
\setcounter{equation}{0}

An example of a unilateral equation was studied by A. Schwarz~\cite{Schwarz}
\be
x^n=1+\epsilon\left(a_0+a_1 x+\ldots+a_{n-1} x^{n-1} \right)\: .
\label{schwarz1}
\ee
For $\eps=0$ we consider $n$ solutions $e^{\frac{2 \pi i k}{n}} 1$, 
$k=0, \ldots, n-1$.
For small $\eps$ we are interested in finding perturbative solutions 
around these.
An explicit iterative expression for $\T \: x^s$ for the solution
$x \stackrel{\eps \rightarrow 0}{\rightarrow} 1$ is~\cite{CeZu}
\be
\T \: x^s =\T \: 1+s \sum_{k=1}^{\infty} \frac{\eps^k}{n^k} 
\! \! \sum_{n_0+ \ldots +n_{n-1}=k} \! \!
\frac{\T {\cal S} (a_0^{n_0} \ldots a_{n-1}^{n_{n-1}})}{n_0!\ldots n_{n-1}!} 
\prod_{r=1}^{k-1} \left(s+\sum_{l=1}^{n-1} l n_l-rn \right)\: .
\label{sol1}
\ee
Some remarks can be made with respect to this formula, before we proceed to
its demonstration.
Its main feature is that it is symmetrized in the $a_r$, which enter 
only through the symmetrized product
$\S(a_0^{n_0} \ldots a_n^{n_n})$. This was conjectured in \cite{ABMZ1} and
subsequently proven in \cite{ABMZ2} and \cite{Schwarz}. 
In (\ref{sol1}) the normalization of the symmetrized product
is chosen in such a way as to give the ordinary product if the factors commute 
\cite{ABMZ2}.

The formula~(\ref{sol1}) holds for positive as well as for negative values 
of $s$.
A similar expression could be derived for all the perturbative solutions of the
equation.

For the equation of the second order $n=2$ an explicit expression can be given
\be
\T \: x=\T \left[ \eps \frac{a_1}{2} + \S
\sqrt{1+\eps a_0+(\eps \frac{a_1}{2})^2} \; \right].
\label{n2}
\ee

Let us now sketch a proof of (\ref{sol1}). We start by applying the Bezout
theorem. For (\ref{schwarz1}) the characteristic polynomial is
\be 
P(\l) \equiv 1-\l^n+\epsilon\left(a_0+a_1 \l+\ldots +a_{n-1} \l^{n-1} \right)
\ee
and
\be
P(\lambda)=Q(\lambda,x)(\lambda-x) \mbox{ for } P(x)=0
\label{bezout2}
\ee
with
\be
Q(\lambda,x)=\sum_{l=0}^{n-1} \left(\sum_{r=l+1}^{n-1} \eps a_r 
x^{r-l-1}-x^{n-l-1} \right) \lambda^l.
\ee
Here, we have chosen one particular solution $x$ of (\ref{sol1}), 
namely the one which satisfies 
$x \stackrel{\eps \rightarrow 0}{\rightarrow} 1$, but the same technique
could be applied to any other of the perturbative solutions.

As a next step, we use the basic property of the trace of the logarithm:
\be
\T \log P(\lambda)=\T \log(Q(\lambda,x))+\T \log(\lambda-x) \: .
\ee
Then, as anticipated in the introduction, we apply Schwarz's idea of making
a contour integration in the complex plane and compute the result 
through the Cauchy theorem.
\be
\T f(x)=(2 \pi i)^{-1} \oint_{\Gamma} d \l \: \T \frac{1}{P(\l)}P'(\l)
f(\l)\: .
\label{integral}
\ee
Here, $\Gamma$ is a small circle around $1$ and $f(\l)$ is a function, which
is regular for $\l$ near $1$. The integration contour is shown in the next
figure.

\begin{figure}[ht]
\centerline{\epsfig{figure=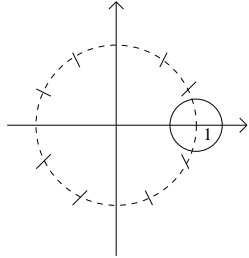,height=0.2\textheight,angle=270}}
\caption{Contour of the integration}
\end{figure}

We factorize
\be
P(\l)=(1-\l^n) T(\l) \: \mbox{ with }
T(\l)=1-\eps (\l^n-1)^{-1} \sum_{l=0}^{n-1} a_l \l^l
\ee
and perform an integration by parts. In this way 
(\ref{integral}) becomes
\be
\T \: f(x)=\T \: f(1) -\frac{1}{2 \pi i} \oint_{\Gamma} d \l \: 
\T \log(T(\l)) f'(\l) \: .
\label{partial}
\ee
We expand the logarithm and make the following change of variable
\be
y=\l^n\: .
\label{transf}
\ee
A small closed curve $\Gamma$ winding once around $1$ still remains 
a closed curve winding once around $1$ after the variable 
transformation (\ref{transf}), so that this is justified.
We restrict ourselves to the case $f(\l)=\l^s$ and get
\be
\T \: x^s=\T \: 1+\frac{s}{n} \sum_{k=1}^{\infty} \frac{\eps^k}{2 \pi i k} 
\oint_{\Gamma} dy \T  
\left(\frac{\sum_{l=0}^{n-1} a_l y^{\frac{l}{n}}}{y-1}\right)^k 
y^{\frac{s}{n}-1}.
\ee
In this form we can already see that the result is symmetrized in
the coefficients $\{a_i\}$, because they enter only through expressions 
of the type
\be
\left(\sum_{l=0}^{n-1} a_l y^{\frac{l}{n}} \right)^k
=\sum_{n_0+\ldots +n_{n-1}=k}\! \! 
\frac{k!}{n_0 ! \ldots n_{n-1}!} \S(a_0^{n_0} \ldots a_{n-1}^{n_{n-1}})
y^{\displaystyle \frac{1}{n}\sum_{l=1}^{n-1} l n_l},
\ee
which are automatically symmetrized.

Finally, (\ref{sol1}) is obtained by applying the Cauchy theorem in its 
more general form
\be
(2 \pi i)^{-1} \oint_C dy \frac{f(y)}{(y-y_0)^k}
=\frac{1}{k-1!} \frac{d^{k-1}}{dy^{k-1}} f(y) |_{y=y_0}\: ,
\ee
where $C$ a closed curve winding once around $y_0$, and $f(y)$ is
a function which is regular inside $C$.

Let us conclude this section by giving some alternative expressions for 
(\ref{sol1}).
We introduce the notation
\be
\left(
\begin{array}{l}
\alpha \\ k
\end{array}
\right)
=\left\{ 
\begin{array}{ll}
\prod_{r=1}^k \frac{\alpha-r+1}{r} & \mbox{ for } k=1,2,\ldots 
\\ \\
1 & \mbox{ for } k=0.
\end{array}
\right.
\ee
Then
\be
\T \: x^s=\T \: 1+
\frac{s}{n} \sum_{k=1}^{\infty} \eps^k (k-1)!
\! \! \! \! \! \! \! \! \sum_{n_0+ \ldots +n_{n-1}=k} \! \! \! \! \! \!\! \!
\frac{\T {\cal S} (a_0^{n_0} \ldots a_{n-1}^{n_{n-1}})}{n_0!\ldots n_{n-1}!} 
\left(
\begin{array}{l}
\frac{1}{n}(s-n+\sum_{l=1}^{n-1} l n_l)\\ \\
k-1
\end{array}
\right).
\ee 
In this form the result can be more easily compared to \cite{ABMZ2}.

Another expression can be given through generalizations of factorials:
\ba
\lefteqn{\T \: x=\T \: 1+\sum_{k=1}^{\infty} \frac{\eps^k}{n^k} 
\! \! \sum_{n_0+ \ldots +n_{n-1}=k} \! \! 
(-1)^{1+[\frac{1}{n}(\sum_{l=0}^{n-1}(l+1)n_{n-1-l}-n-1)]}} \\
&& \! \! \! \! \! \!\! \!
N^{(n)}\left(\sum_{l=0}^{n-1}(l+1)n_{n-1-l}-n-1\right) 
N^{(n)}\left(\sum_{l=1}^{n-1}l n_l-n+1\right)
\frac{\T {\cal S} (a_0^{n_0} \ldots a_{n-1}^{n_{n-1}})}{n_0!\ldots n_{n-1}!}, 
\nonumber
\ea
where $[x]=$ Integer Part of $x$ and
\ba
\displaystyle{N^{(n)}(x)=}\left\{
\begin{array}{ll}
1 & \mbox{ for } x<0 \\ \\
\prod_{r=0}^{[x/n]} x-rn
& \mbox{ for } x \ge 0.
\end{array}
\right.
\ea
This formula can be more easily compared with the exact expression (\ref{n2})
for $n=2$, as the symbols $N^{(n)}(x)$ are generalizations of the double 
factorials, which appear in the expansion of the square root.

\section{Discussion of another equation}
\setcounter{equation}{0}

Another unilateral matrix equation is studied in \cite{ABMZ2}:
\be
\Phi=A_0 +A_1 \Phi+ \ldots A_n \Phi^n\: .
\label{ABMZ}
\ee

Its characteristic polynomial is 
\be
\l-A(\l) \mbox{ where } A(\l)=A_0 +A_1 \l+ \ldots A_n \l^n.
\ee
We follow a similar procedure as before: we apply the 
Bezout theorem and perform a contour integration. Then
\be
\T f(\Phi)=-\frac{1}{2 \pi i} \oint_C d \l \: \T \log(1-\frac{A(\l)}{\l}) 
f'(\l),
\ee
where now $C$ is a closed curve winding once around $0$.
We restrict to the case $f(\l)=\l^s$, $s$ positive integer, 
expand the logarithm and apply the Cauchy theorem: 
\be
\T \: \Phi^s=s \T \sum_{k=1}^{\infty} \frac{1}{k} (A_0+\ldots A_n)^k 
|_{\sum_{l=0}^n (l-1)n_l=-s}.
\label{Dan}
\ee
In this way we recover the result of \cite{ABMZ2}.

To study more closely the relation between equation~(\ref{schwarz1}) 
and (\ref{ABMZ}) we make the Ansatz
\be
x=1+\alpha \Phi\: \mbox{ with } \alpha^{n-1}=-n\: .
\label{choice}
\ee
Then it is easily seen that
\be
A_l= \left\{
\begin{array}{ll}
 -\alpha^{l-n} \eps \displaystyle{\sum_{r=l}^{n-1}}
\left(\begin{array}{l} r\\ l \end{array} \right)
a_r & \mbox{ for } l=0,1 \\
\alpha^{l-n} \left(\begin{array}{l} n\\ l \end{array} \right)
-\alpha^{l-n} \eps \displaystyle{\sum_{r=l}^{n-1}}
\left(\begin{array}{l} r\\ l \end{array} \right)
a_r & \mbox{ for } 2 \le l \le n\: .
\end{array}
\right.
\label{transform}
\ee
Some remarks can be made with respect to (\ref{Dan}) and (\ref{transform}).

For $\eps=0$ the solution of (\ref{ABMZ}) is $\Phi=0$. Therefore it has 
no sense to invert 
$\Phi$ and $s$ has to be a positive integer. However, negative powers of 
$x$ can still be computed by expanding (\ref{choice}) into a series of 
positive powers of $\Phi$.

The equation (\ref{ABMZ}) depends on $n+1$ coefficients $A_r$, but there are 
only $n$ coefficients $a_r$ in (\ref{schwarz1}). 
Therefore in (\ref{transform}), $A_n=1$ is fixed.

Since most of the solutions of the Schwarz's equations are not real, 
even for real $a_n$, there is no reason to choose $\alpha$ to be real in 
(\ref{choice}).

The transformation (\ref{transform}) is linear and invertible.
As a  consequence, the result that the series is symmetrized needs to 
be proven only for one the two equations, and then it follows immediately
for the other.
However, from (\ref{transform}) it is clear that even if $\eps$ is small, only 
the first two coefficients $A_0$ and $A_1$ need to be small. 
We do not study the convergence properties of the sums appearing in this
note, and we consider them always as formal series, but it is to be
expected that the expansions do not hold for the same range of the 
coefficients. 

\section*{Acknowledgments} 

We would like to thank Dan Brace and Bogdan Morariu for many helpful
discussions.
This work was supported in part by the Director, Office of Science,
Office of High Energy and Nuclear Physics, Division of High Energy Physics of 
the U.S. Department of Energy under Contract DE-AC03-76SF00098 and 
in part by the
National Science Foundation under grant PHY-95-14797. B.L.C. is supported
by the DFG (Deutsche Forschungsgemeinschaft) under grant \mbox{CE 50/1-1}. 
B.L.C. would like to thank the organizers of the euroconference for the
possibility of attending it and all the participants for the nice atmosphere.

\end{document}